\documentstyle[epsf,12pt]{article}

\catcode`\@=11
\@addtoreset{equation}{section}
\catcode`\@=12
\begin{document}
\baselineskip8mm
\title{\vspace{-6mm}Topological entropy for some isotropic
cosmological models}
\author{A. Yu. Kamenshchik$^{1,2}$, \ I. M. Khalatnikov$^{1,2,3}$
 \ S. V. Savchenko$^{1}$\\  and \ A. V. Toporensky$^{4}$}
\date{}
\maketitle
\hspace{-1mm}$^{1}${\em L.D. Landau Institute for
Theoretical Physics, Russian
Academy of Sciences, Kosygin str. 2, Moscow, 117334, Russia}\\
$^{2}${\em Landau Network Centro Volta, Villa Olmo, Via Cantoni 1,
Como, 22100, Italy}\\
$^{3}${\em Tel Aviv University,
Tel Aviv University, Raymond and Sackler
Faculty of Exact Sciences, School of Physics and Astronomy,
 Ramat Aviv, 69978, Israel}\\
$^{4}${\em Sternberg
Astronomical Institute, Moscow University, Moscow, 119899, Russia}\\
\begin{abstract}

The chaotical dynamics is studied in different Friedmann-Robertson-
Walker cosmological models with scalar (inflaton) field and
hydrodynamical matter. The topological entropy is calculated for some
particular cases.  Suggested scheme can be easily generalized for wide
class of models.  Different methods of calculation of topological
entropy are compared.
\end {abstract}

 \section{INTRODUCTION}
 In recent years the
problem of chaos in general relativity and classical cosmology has
attracted  great attention \cite{chaos}.  A special efforts has been
made to resolve \cite{BKL-recent} the questions concerning the chaotical
nature of the oscillatory approach to singularity \cite{BKL} or in
other terms the chaotic nature of the mixmaster universe
\cite{Mixmaster}. However, already simple isotropical closed
Friedmann-Robertson-Walker models manifest some elements of chaotical
behavior which should be taken into account for a correct construction of
quantum cosmological theories  \cite{Corn-Shel}.

The study of classical dynamics of closed isotropical cosmological
model has a long history. First, it was noticed that in such models
with a minimally coupled massive scalar field there is an opportunity to
escape a singularity at contraction \cite{Ful-Park-Star}. Then the periodical
trajectories escaping a singularity were studied \cite{Hawking}.
In Ref. \cite{Page} it was shown that the set of infinitely bouncing
aperiodical trajectories has a fractal nature.  Later this result in
other terms was reproduced in our papers \cite{we,we1,we2}.

Here we would like
to describe briefly the approach presented in \cite{we}. The main idea
consisted of the fact that in the closed isotropical model with a
minimally coupled massive scalar field all the trajectories have the
point of maximal expansion. The localization of the points of maximal expansion
on the configuration plane $(a,\phi)$, where $a$ is a cosmological radius,
while $\phi$ is a scalar field, could be found analytically. Then
the trajectories could be classified according to localization of their
points of maximal expansion. The area of the points of maximal expansion
is located inside the so-called Euclidean or ``classically forbidden''
region. Numerical investigation shows that this area has a
quasiperiodical structure, zones corresponding to the falling to singularity
are intermingled with zones in which are placed points
of maximal expansion of trajectories having the so-called ``bounce''
or point of minimal contraction. Then studying the substructure of
these zones from the point of view of the possibility of having two bounces,
one can see that this substructure repeats on the qualitative level the
structure of the whole region of the possible points of maximal
expansion.  Continuing this procedure {\it ad infinitum} one can see
that as the result one has the fractal set of infinitely bouncing
trajectories.

The same scheme gives us an opportunity to
see that there is also a set of periodical trajectories.
All these periodical trajectories contains bounces intermingled with a
series of oscillations of the value of the scalar field $\phi$.
It is important that there are no restrictions on the lengths of series
of oscillations in this case.
In the paper of Cornish and Shellard
\cite{Corn-Shel}  the topological entropy was calculated for this case
by the methods of symbolic
dynamics
and it was shown that it is positive.
Using
symbolic dynamics
is not new
for models
of theoretical physics.
For example,
in the well-known paper
\cite{Gaspard and Rise}
it is used for
investigation of the chaotic
repellor, generated by the scattering of a point
particle from three hard discs in a plane,
which one can consider
as a model of
the time evolution
of a metastable
configuration of particles.
This repellor
consists of
the trajectories
which remain confined to the scattering
region as $t\to \infty.$
If three hard discs with radius $a$
are fixed in the plane
at the vertices of an equilateral
triangle with side $R$
and $R > 3a,$
then all the trajectories
of this repellor are
in one-to-one
correspondence
with doubly
infinite
sequences $...x_{-2}x_{-1}x_{0}x_{1}x_{2}...$
of the symbols $\{1,2,3\}$
with the natural
constraint $x_{n+1}\neq x_{n}.$
The symbolic dynamics
for it is very simple:
in the sequences
only doublets
such as $11,$
$22$ and $33$
do not occur.
The topological
entropy of the system
equals $\ln 2$
and therefore
such a repellor is chaotic.

Below we reproduce the
calculations of Ref. \cite{Corn-Shel} and show how it is possible to
generalize it for more complicated cases, but first we describe
how introduction of matter or another modification of the model changes
the structure of periodical trajectories. The structure of the paper
is the following: in the second section we describe the
properties  of cosmological models under investigation, while in the
third section we present the algorithm for the calculation of
the topological entropy.

\section{COSMOLOGICAL MODELS}
First let us write down the action for the simplest cosmological
model with the scalar field:
\begin{equation}
S = \int d^{4} x \sqrt{-g}\left\{\frac{m_{P}^{2}}{16\pi} (R -
2\Lambda) + \frac{1}{2} g^{\mu\nu}\partial_{\mu}\phi
\partial_{\nu}\phi -\frac{1}{2}m^{2}\phi^{2}\right\},
\label{action}
\end{equation}
where $m_{P}$ is the Planck mass, $\Lambda$ is the cosmological constant.
The equations of motion for a closed isotropic universe are
\begin{equation}
\frac{m_{P}^{2}}{16 \pi}\left(\ddot{a} + \frac{\dot{a}^{2}}{2 a}
+ \frac{1}{2 a} \right)
+\frac{a \dot{\phi}^{2}}{8}
-\frac{m^{2} \phi^{2} a}{8}-\frac{m_{P}^{2}}{8 \pi}\Lambda a = 0,
\label{equation1}
\end{equation}
\begin{equation}
\ddot{\phi} + \frac{3 \dot{\phi} \dot{a}}{a}
+ m^{2} \phi = 0.
\label{equation2}
\end{equation}
The first integral of motion of our system is
\begin{equation}
-\frac{3}{8 \pi} m_{P}^{2} (\dot{a}^{2} + 1)
+\frac{a^{2}}{2}\left(\dot{\phi}^{2} + m^{2} \phi^{2} +
\frac{m_{P}^{2}}{8 \pi} \Lambda \right)  =
0.
\label{integral}
\end{equation}
In the case when the cosmological constant is equal to zero, the form of the
boundary of the
Euclidean region is given by an equation which can be easily obtained
from Eq. (\ref{integral}):
\begin{equation}
m^{2} a^{2} \phi^{2} = \frac{3}{4\pi}m_{P}^{2}.
\label{Euclid}
\end{equation}
The form of this Euclidean region is shown in Fig. 1(a).
In this model, investigated in many papers
\cite{Hawking,Page,we,Corn-Shel} , there are periodical trajectories
with an arbitrary number of oscillations of the scalar field.

It can be understood by using two famous asymptotics for this dynamical system.
For large $\phi$ the slow-roll regime
leads to a quasiexponential growth of the scale
factor (so-called inflation), and in the end of this regime
$$
a_{end}=a_0 exp(2 \pi (\phi_0/M_p)^2)
$$
where $a_0$, $\phi_0$ are the initial values and
 $a_{end}$ is the scale factor in the end
of inflation.

For small $\phi$ the massive scalar field looks like  dust perfect
fluid but with small oscillations with the frequency equal to $m$.
Using the Friedmann solution for a closed dustlike universe, it is possible
to express the scale factor in the point of maximal expansion $a_{me}$ through
the value of the scale factor at the end of inflation $a_{end}$:

$$
(m a_{end})^3 \sim m a_{me}
$$
and the duration of the Friedmann stage

$$
T= \pi a_{me}.
$$

So, starting with $\phi_{0}$, $\phi_{0} >> m_{P}$, the universe will go through
$$
N=m^3 a_{end}^3 = m^3 a_0^3 exp(6 \pi
(\phi_0/M_p)^2)
$$
oscillations. This value can be arbitrarily large,
because the value of the scalar
field at the Euclidean boundary can also be arbitrarily large.

Of course, it in necessary to prove that the bounce is indeed possible
for any large value $a_{me}$. It was done analytically by Starobinsky in
\cite{Ful-Park-Star} with the estimation of the bounce
probability depending on $a_{me}$.

Inclusion of the positive cosmological constant investigated in \cite{we1}
implies two opportunities: if $\Lambda$ is small in comparison with
the mass square of the scalar field $m^2$,
 the qualitative behavior is the same as in the
model without cosmological constant; if the cosmological constant is of order
$m^{2}$ the chaotical dynamics disappears in a jumplike manner \cite{we1}.

It is interesting from a calculational point of view
to consider the case of the negative cosmological constant. In this case the
Euclidean region has the form presented in Fig. 1(b) and numerical calculation
shows that the possible number of oscillations is restricted. With the
growth of the absolute value of the cosmological constant this number
is decreasing and at $|\Lambda| \sim 0.34 m^2$ all the periodical
trajectories disappear and dynamics become regular.

Let us include into consideration hydrodynamical matter with the
equation of state
\begin{equation}
p = \gamma \epsilon,
\label{state}
\end{equation}
where $p$ is the pressure, $\epsilon$ is an energy density, and $\gamma$ is
the constant ($\gamma = 0$ corresponds to dust matter, $\gamma = 1/3$
corresponds to radiation, while $\gamma = 1$ describes the massless
scalar field). Our analysis is valid for $\gamma > -1/3$.
In this case the form of the boundary of the Euclidean region is given
by equation
\begin{equation}
m^{2} a^{2} \phi^{2} = \frac{3}{4\pi}m_{P}^{2} - \frac{2 D}{a^{q}},
\label{Euclid1}
\end{equation}
where $D$ is the constant characterizing the quantity of the given type
of matter in the universe and $q = 3(\gamma+1) - 2$. The form of the Euclidean
region described by Eq. (\ref{Euclid1}) is shown in Fig. 1(c). One can
see that in this case the Euclidean region is restricted from above.
Numerical calculation shows that in this case again only a restricted number
of oscillations is possible. Moreover, the structure of periodical
trajectories in this case is much more complicated. Indeed, the law is the
following long series of oscillations that can come after the bounce, followed by a
short series of oscillations, while the behavior of the trajectory after a
short series of oscillations is less restricted.
The concrete laws governing the structure of trajectories depend on
the parameters of the model under consideration.
However, in this case
also it is possible to calculate a topological entropy, which will be
demonstrated below.  Completing a discussion about the model with
hydrodynamical matter one has to say that a large amount of matter
suppresses chaotic behavior and at some critical values of the constant
$D$ periodical trajectories disappear and all the trajectories go to
the singularity.  For the above-mentioned cases, interesting from a
physical point of view, values are the following: for the
case $q = 4$ (massless scalar field), one has $$D m^4 > 0.0028
m_{P}^{2};$$ for $q = 2$ (radiation), $$D m^2 > 0.0093 m_{P}^{2};$$ and
for $q = 1$ (dust matter), $$D m > 0.023 m_{P}^{2}.$$

It is necessary to underline that bounces are still possible in such a
situation (unless $D$ is much bigger than a critical value), but there are
no periodical trajectories, because immediately  after bounce the trajectory
has a point of maximal expansion and restores their travel to
singularity.

Now we would like to consider the last example: the cosmological model
with a complex scalar field which was studied earlier in papers
\cite{complex}.
Using  the most natural representation of the complex scalar field
\begin{equation}
\phi = x \exp(i\theta),
\label{complex}
\end{equation}
where $x$ is an absolute value of complex scalar field while
$\theta$ is its phase. This phase is cyclical variable corresponding to
the conserved quantity - the classical charge of the universe, which
plays the role of the quasifundamental constant of the theory:
\begin{equation}
Q = a^{3} x^{2} \dot{\theta}.
\label{charge}
\end{equation}
The shape of the Euclidean region presented in Fig. 1(d) is given now by
\begin{equation}
m^{2} a^{2} x^{2} = \frac{3}{4\pi}m_{P}^{2} - \frac{Q^{2}}{a^{4} x^{2}}.
\label{Euclid2}
\end{equation}
Again, as in the preceding example at some critical value of
the charge $Q$ periodical trajectories and chaotical dynamics
vanish, while at lower values of charge there are again complicated rules
governing the structure of periodical trajectories.
The condition of regular behavior  is given by the inequality:
$$Qm^{2} > 0.056 m_{P}^{2}.$$

\begin{figure}
\epsfxsize=\hsize
\centerline{\epsfbox{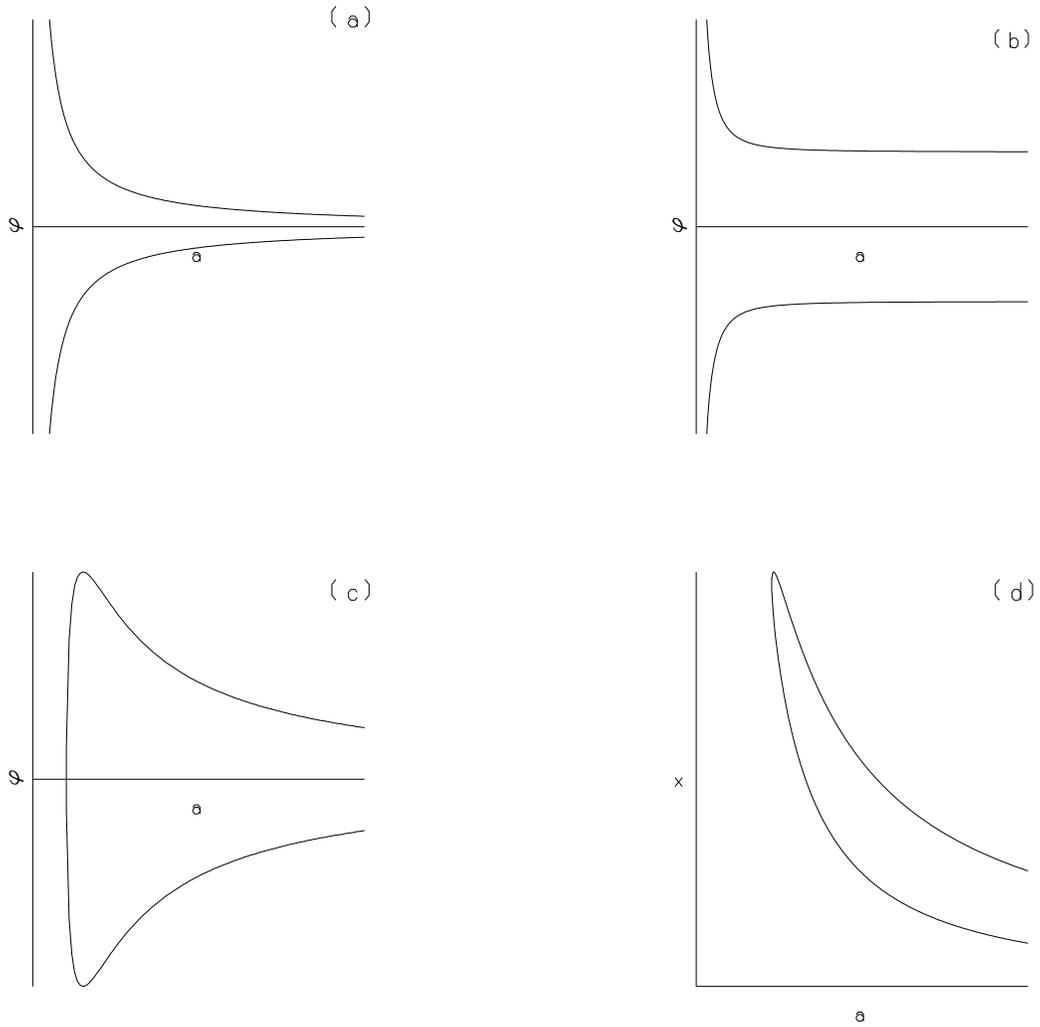}}
\caption{In Fig. 1
the shapes of Euclidean regions for different isotropic cosmological
models with the scalar field are presented;
a) simplest model with the scalar field,
b) model with negative cosmological constant,
c) model with hydrodynamical matter,
d) model with complex scalar field.}
\end{figure}

It is important to notice that the real structure of trajectories
in the case of isotropic models with scalar field and matter or with
complex scalar field is more complicated because there are two types of
bounces. These bounces could be called ``upper'' and ``lower''
depending on the value of the scalar field at which they take place. The
value of the scalar field separating the two types of bounces correspond to
a periodic trajectory with a maximal number of oscillations of the scalar
field. However, in what follows we shall not discriminate between the two
types. Generalization of suggested algorithms for the calculation of
topological entropy for the ``alphabet'' including two types of bounces
is straightforward.

\section{CALCULATION OF THE TOPOLOGICAL ENTROPY}
Now, let us calculate the topological entropy for some of the considered
cases.
In our situation
strange repellor $\Omega$ is the set of all
perpetually bouncing universes.
We shall find
topological entropy
of the Poincare return map $S$ with discrete time
(see Ref. \cite{Arnold-Avez}), generated
by the surface of section $X$
consisting of the points,
in which bounce or oscillation occurs:
if $x$ is the point of
intersection (crossing)
of the trajectory with
the surface $X,$
then $Sx$ is
the point of its following
intersection with $X$
(by definition  of $\Omega$
each trajectory
of $\Omega$ crosses
$X$ infinitely
many times).
Such an entropy is independent
of phase-space coordinates.
We note that the topological entropy depends on
the choice of Poincare surface for the section.
Our
Poincare surface of section $X$
is
natural for cosmological models with a
massive scalar field for
closed Friedmann-Robertson-Walker universes
and plays the same role as
the boundary of discs or
obstacles for billiard systems.

Topological entropy $H_{T}$ "measures" the
chaos of the dynamical system.
There are many equivalent
definitions of topological entropy
(via finite partition of
the phase space) in mathematics (see Ref. \cite{Walt}).
But for many
dynamical systems (in particular,
for the finite topological
Markov chain, to which
the symbolic systems considered
below
belong)
this quantity
coincides with
the exponent of
growth in the number of periodic
points as their period increases.
We remark that the set of periodic points
and the set of all closed orbits
are
everywhere
dense in space $X$
and in the repellor $\Omega,$
respectively, and
thus, they contain
basic information
about
dynamics.

More precisely, let
$(S,X)$ be a
dynamical system with
discrete time.
If $x \in X$ and
$S^{k}x=x,$
the point $x$
is called a periodic point of period $k.$
We denote by
$N(k)$ the number of periodic points of period or length $k$.  Then, the
definition of topological entropy is as follows:
\begin{equation}
H_{T} = \limsup_{k \rightarrow \infty} \frac{1}{k} \ln N(k).
\label{define}
\end{equation}
If $H_{T} > 0$, one can conclude that the dynamics is chaotic.

One can quantify the length of the orbit
by the number of symbols.
To begin with let us reproduce in some detail calculations from
Ref. \cite{Corn-Shel}. It was a discrete coding of orbits, using two
symbols:\\
$A$, which is the bounce of the trajectory, and\\
$B$, which is the crossing of line $\phi = 0$.

It is important that
for our repellor
for a given admissible infinite sequence
of symbols $A$ and $B$ there is only one
trajectory, which goes through bounces and oscillations in the order
defined by this sequence.  Therefore, in this sequel we can deal with only
such sequences.

For the simplest model with the scalar field there is the only
prohibition rule: two letters $A$ cannot stay together, which means
that it is impossible to have two bounces one after another without
oscillations between them (this condition
is analogous to the impossibility
of having two successive collisions
about the same disc for the scattering from three discs).
Thus our symbolic system
is a topological Markov chain (of the first order)
with two letters (see corresponding definition  in Ref.
\cite{Parry-Poll}).  Each topological Markov chain  of the first order
with $k$ letters $\{B_{1},...,B_{k}\}$
is defined by
topological transition matrix $D.$
It is a $k\times k$ matrix of zeros and ones,
the set of indices
of which coincides
with the set of letters.
The entry
$D_{B_{1}B_{2}}$  of the matrix $D$
is zero
precisely when $B_{1}B_{2}$
is a prohibited word of length 2.  In our case, the topological
transition matrix $D$ is
$$
\left(
\begin{array}{cc}
 0 & 1\\ 1 & 1
\end{array}
\right)
$$
 (the zero in this matrix corresponds to the prohibition
of the word $AA$).  There are many practical means of counting the
topological entropy.  For example, calculating the trace of the matrix
$D^{k},$  we obtain the number of all period $k$ points of the map $S.$
Since $Tr\ D^{k}=\lambda_{1}^{k}+\lambda_{2}^{k},$
where $\lambda_{1}$ and $\lambda_{2}$ are
eigenvalues of the matrix $D,$
it is sufficient
to find the spectrum of $D.$
In fact, we must calculate
only the largest positive eigenvalue, which
exists by the Perron-Frobenius theorem
(see Ref. \cite{Gant};
such an eigenvalue is called
Perron-Frobenius's number).
In our case $\lambda_{1}=\lambda_{2}^{-1}=(\sqrt{5}+1)/2.$
The logarithm of this number is
sought to define the topological entropy.
Another means of calculating
the topological entropy consists of
finding the smallest
positive root of the equation
$\varphi_{A}(z)=1,$
where $\varphi_{A}(z)$ is a generating function
of periodic points without intermediate
recurrences in the initial symbol $A$
(see Ref. \cite{Peter})
In our case the next
periodic points are of the following type:
$ABA,$ $ABBA,...,ABBBBB...BBA,....\ .$
Thus $\varphi_{A}(z)=z^{2}+z^{3}+...+...=z^{2}/(1-z).$
The smallest positive root of the equation
$z^{2}/(1-z)=1$ is $(\sqrt{5}-1)/2.$
The logarithm of the reciprocal
from
this number
is again
a topological entropy.

Now we present
the method of recurrent relations,
which as one can see in sequel
is more successful for our
purposes.
Let us denote by $Q(k)$ the number of ``words'' (trajectories) of length
 $k$ satisfying this rule, which begin with $A$ and end with $A$, and
by $P(k)$ the number of words which begins with $A$ and end with $B$.
(We remark that
our definition of
coefficients $Q(k)$ and $P(k)$
is distinguished from that of  Ref.
\cite{Corn-Shel}: in their paper
$P(k)$ is the number
of period k points ending in A,
and $Q(k)$ is the number
of period k points ending in B).
Then one can easily write down the recurrent relations:
\begin{eqnarray}
&&Q(k+1) = P(k),\nonumber \\
&&P(k+1) = Q(k) + P(k).
\label{recurr}
\end{eqnarray}
It is easy to get from Eqs. (\ref{recurr})
the following relation for $P(k)$:
\begin{equation}
P(k+1) = P(k) + P(k-1).
\label{recurr1}
\end{equation}
One can easily calculate that
\begin{equation}
P(2) = 1, \qquad P(3) = 1
\label{initial}
\end{equation}
and that Eq. (\ref{recurr1}) defines the series of Fibonacci numbers.
Let us recall how to find the formula for the general term
of the Fibonacci series. One can  look for $P(k)$ as a linear
combination of terms $\lambda^{k}$, where $\lambda$ is the solution
of equation
$$
\lambda^{k+1} = \lambda^{k} + \lambda^{k-1}
$$
or, equivalently, because we are interested only in nonzero roots
\begin{equation}
\lambda^{2} - \lambda - 1 = 0.
\label{equatio}
\end{equation}
Looking for $P(k)$ in the form
\begin{equation}
P(k) = c_{1} \lambda_{1}^{k} + c_{2}\lambda_{2}^{k},
\label{form}
\end{equation}
where $\lambda_{1}$ and $\lambda_{2}$ are the roots of Eq.
(\ref{equatio})
and satisfying the conditions (\ref{initial}) one can get
\begin{equation}
P(k) = \frac{1}{\sqrt{5}}\left[\left(
\frac{1+\sqrt{5}}{2}\right)^{k-1} + (-1)^{k-2}
\left(\frac{\sqrt{5}-1}{2}\right)^{k-1}\right].
\label{solution}
\end{equation}
Substituting Eq.(\ref{solution}) in the definition of the topological
entropy (\ref{define}) one can find
\begin{equation}
H_{T} = \ln \left(\frac{1+\sqrt{5}}{2}\right) > 0,
\label{entropy}
\end{equation}
where
$\left(\frac{1+\sqrt{5}}{2}\right)$
is the famous golden mean. It is clear that only the largest root
of Eq. (\ref{equatio}) is essential for the calculation of
the topological entropy.

We would like to point out
that
the number $N(k)$
of all
period k points
grows as $\exp(H_{T}k)$
for $k \to \infty.$
In our case the number
$Q(k+1)$ is in fact
the number
of all real
closed orbits (with respect to
dynamical  system with continuous time,
generated by Einstein's equations),
which are coded by $k$ symbols
(here we distinguishes
closed orbit and it's multiple).
In other words,
$Q(k+1)$ is the number
of all periodic orbits
of period $k$
(each periodic point $x$ of period $k$
generates periodic
orbit $\{x,Sx,...,S^{k-1}x\};$
it is clear that points $Sx,...,S^{k-1}x$
generate  the same periodic orbit; thus
the number of all period $k$ points
is much greater than that of all
periodic orbits of period
$k$)
and its multiples.
This number
grows as $\mu(\{x_{0}=A\})\exp(H_{T}k)$
for $k \to \infty,$
where
$\mu$ is the measure
of maximal entropy
for our topological
Markov chain
and $\{x_{0}=A\}$ is the set
of all sequences,
which start with the letter $A$
(so-called "one-dimensional" cylindrical set) \cite{Bow}.
Such a measure
is characterized
by the next important
property: it
is the most chaotic
measure among
all invariant
probability
measures for
our topological Markov chain
(more precisely,
it is a measure for which
the Kolmogorov-Sinai
entropy is equal to
the topological entropy).
For example,
for the set of
all infinite sequences
of the letters $A$ and $B$,
such a measure is the
Bernoulli measure $\nu$
with
$\nu(\{x_{0}=A\})=\nu(\{x_{0}=B\})=1/2.$
In the case of
a topological Markov chain
it is
a Markov measure.
A measure with maximal entropy
plays an important
role in the distribution of
periodic points.
There are many
means of counting
the probability
of "one (and multi)-dimensional"
cylindrical sets
for such a measure.
In particular,
one can find
the positive eigenvectors $\xi$ and $\xi^{*}$
for Perron-Frobenius's
eigenvalue of the topological transition
matrix $A$ and its conjugate $A^{*},$
respectively.
Taking
the product of the corresponding
components of the vectors $\xi$ and $\xi^{*}$
and then normalizing it,
we obtain  the probability
of each "one-dimensional"
cylindrical set.
On the other hand,
this number is
coefficient
before
the $k$ power of the
largest
eigenvalue
in the  decomposition
of $Q(k+1)=P(k).$
Thus, in our case it is equal
to $c_{1}=(\sqrt{5}-1)/2\sqrt{5}.$

Now one can go to a more involved case of the cosmological model
with scalar field and negative cosmological constant. As was described
above, in this model the periodical trajectories can have only a
restricted number of oscillations of the scalar field $\phi$. This rule
can be encoded in the prohibition to have more that $n$ letters $B$
staying together,
where the number $n$ depends on the parameters of this model (this
dependence obtained by numerical calculations is represented in Fig.
2).

\begin{figure}
\epsfxsize=\hsize
\centerline{\epsfbox{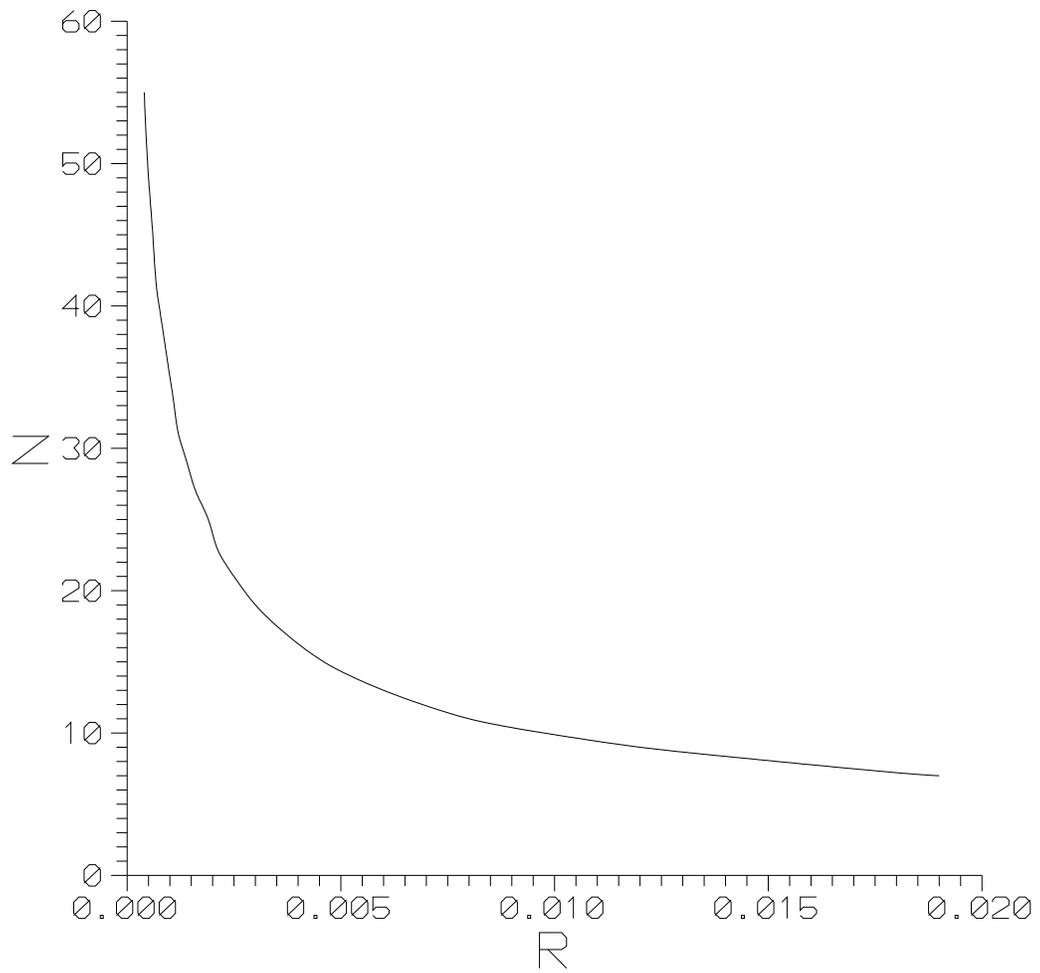}}
\caption{In Fig. 2 the dependence of the number of possible
oscillations $N$ on the ratio $R=|\Lambda|/m^{2}$ is represented.}
\end{figure}

Thus we have a topological Markov chain
of order $n+1.$
Correspondingly, the order of the
topological transition matrix $D(n)$
is equal to $2^{n+1}$.
We see that the first method
is not  well applicable to our
case (in general, to any typical big matrix;
this method is useful
for theoretical
investigations).
Using the second method
we have
$\varphi_{A}(z)=\sum\limits_{k=0}^{n-1}z^{2+k}=z^{2}(z^{n}-1)/(z-1)$
and therefore topological
entropy is equal to
$-\ln z(n),$ where
$z(n)$ is the smallest root of the next equation
$z^{n+2}-z^{2}-z+1=0$
(the largest positive
eigenvalue $\lambda(n)$
of the matrix
$D(n)$ is connected
with the number $z(n)$
by the next relation:
$\lambda(n)=z(n)^{-1}$).

At this time the recurrent relations is as follows:
\begin{eqnarray}
&&Q(k+1) = P(k),\nonumber \\
&&P(k+1) = Q(k) + P(k) - Q(k-n)\theta(k-n),
\label{recur}
\end{eqnarray}
where the $\theta$ - function is defined in the usual manner.
We are interested in the limit $k \rightarrow \infty$ and can
substitute instead of $\theta(k-n)$ number $1$. Now one can write down
the recurrent relation:
\begin{equation}
 P(k+1) = P(k) + P(k-1) -
P(k-n-1),
\label{recur1}
\end{equation}
which in turn implies the
following equation for topological entropy:
\begin{equation}
\lambda(n)^{n+2} - \lambda(n)^{n+1} - \lambda(n)^{n} + 1 = 0,
\label{equatio1}
\end{equation}
where topological entropy is equal to the logarithm of the biggest root
of Eq. (\ref{equatio1}):
\begin{equation}
H_{T} = \ln \lambda(n).
\label{entropy1}
\end{equation}
For the small values of $n$, the biggest root of Eq. (\ref{equatio1})
could be found analytically.

For $n = 1$, we have $\lambda(1) = 1$
and topological entropy is equal to zero
and chaotic behavior is absent, which is clear from physical point of
view.

 For $n = 2$,  we obtain
\begin{equation}
 \lambda(2) =
\frac{1}{3}\left(\frac{27}{2} - \frac{3\sqrt{69}}{2}\right) ^{1/3}
+\frac{\left(\frac{1}{2}(9+\sqrt{69})\right)^{1/2}}{3^{2/3}} \approx
1.32 .
 \label{m2}
\end{equation}
 For $n = 3$, one can get
\begin{eqnarray}
&&\lambda(3) = \frac{1}{3} +
\frac{1}{3}\left(\frac{29}{2}-\frac{3\sqrt{93}}{2}\right)^{1/3}\nonumber \\
&&+\frac{1}{3}\left(\frac{29}{2}+\frac{3\sqrt{93}}{2}\right)^{1/2}
\approx 1.47 .
\label{m3}
\end{eqnarray}
For higher values of $n$ one can find $\lambda$ numerically,
for example for $n = 4$, $\lambda(4) \approx 1.53$.

For large values of $n$ one can find an asymptotical value for
the biggest root $\lambda$:
\begin{equation}
\lambda(n) = \frac{1 + \sqrt{5}}{2} - \frac{1}{2}
\left(\frac{\sqrt{5}-1}{2}\right)^{n-1}+O\left(\left(
\frac{\sqrt{5}-1}{2}\right)^{2n}\right).
\label{asymp}
\end{equation}

Thus $\lambda(n)$
converges to golden mean
exponentially fast when $n\to \infty$
(or, negative cosmological
constant tends to zero).

Concerning the coefficient
before $k$-power of
the largest eigenvalue
or, the probability   of
"one-dimensional" cylindrical
set corresponding to the letter $A$
we may show that in our case
it is equal to the reciprocal of
$z\frac{d}{dz}\varphi_{A}(z)|_{z=z(n)}=2+
\frac{z(n)}{1-z(n)}-n\frac{z(n)^{n+2}}{1-z(n)}$
(see Ref. \cite{Peter}).
Since $z(n)$ tends to $(\sqrt{5}-1)/2$
exponentially fast when $n\to \infty,$
we see that
it is true for the
convergence of $\mu_{n}(\{x_{0}=A\})$
to $\mu(\{x_{0}=A\}),$
where $\mu_{n}$ is a
measure of maximal entropy
for the topological Markov chain
with the topological transition matrix $D(n).$
Thus
two basic parameters
in the distribution
of periodic orbits
of our approximating
systems
develop very regular
behavior when the
negative
cosmological constant
tends to zero.

As it has already been described above in the model with the scalar
field and matter or in the model with the complex scalar field and
nonzero classical charge, the rules governing the structure are rather
complicated.  Nevertheless, in this case also it is possible to
calculate the topological entropy. Here we shall consider one
particular example, however, the algorithm which shall be presented
could be used for different sets of rules as well.

Thus, let us formulate rules for our model:
(1) It is impossible to have more than 19 letters $B$;
(2) after a series with 19 letters $B$ (and letter $A$), one can have
the next series only with 1 letter $B$;
(3) after a series with 18 letters $B$, one can have the next series
with 1 or 2 letters $B$;
(4) after a series with 17 letters $B$, one can have the next series
with 1, 2 or 3 letters $B$, \ldots  (20) after a series with 1 letter
$B$,  one can have the series with $n$ letters $B$, where $0 \leq n
\leq 19$.

These rules define the topological Markov chain of
order 22. The method of
calculating the topological
entropy via  finding
the spectrum of the
topological transition
matrix and the method of
generating functions
are not applicable to this case.
But the method of recurrent relations
allows us to write the equation for
topological entropy
(we do not calculate
the coefficient before the
$k$ power of
the largest
eigenvalue in the distribution
of periodic orbits
because  it is a very
difficult problem now).

Let us notice that the system of rules has a remarkable symmetry with
respect to the number $n_{C} = 10$, which makes further calculations
more simple. These symmetrical rules give us a good approximation for
the description of a real physical situation. The value $n_{C}$ is
apparently a function of the parameters of the model under
investigation.  More detailed numerical investigation implies a more
complicated system of rules, however, it also could be formalized in
the system of recurrent relations. Below we shall see that already the
symmetric system of rules and a comparatively small number, which we
have chosen for our example ($n_{C} = 10$) gives rather a cumbersome
equation for the topological entropy.

Let us now introduce the following notation:\\
$Q(k)$ is the number of words which begin with letter $A$
and end with letter $A$,\\
$Q_{1}(k)$  is the number of words beginning with letter $A$ and ending
with a series with 1 letter $B$,\\
$Q_{2}(k)$ is the number of words beginning with letter $A$ and ending
with a series with 2 letters $B$,\\
\ldots,\\
$Q_{19}(k)$ is the number of words beginning with letter $A$ and
ending with a series with 19 letters $B$.\\

The system of recurrence relations for these quantities is as
follows:
 \begin{eqnarray}
 &&Q(k+1) = Q_{1}(k) + Q_{2}(k) + \cdots +
Q_{19}(k),\nonumber \\ &&Q_{1}(k) = Q(k-1),\nonumber \\ &&Q_{19}(k) =
Q_{1}(k-20),\nonumber \\ &&Q_{d}(k) = Q_{d-1}(k-1) - Q_{21-d}(k-d-1),\
2 \leq d \leq 10 \nonumber \\ &&Q_{d}(k) = Q_{d+1}(k+1) +
Q_{20-d}(k-d-1),\ 11 \leq d \leq 18
 \label{req}
 \end{eqnarray}
Resolving this system with respect to $Q(k)$, one get the following
recurrent relation:
\begin{eqnarray}
&&Q(k+1) = Q(k-1) + Q(k-2) + Q(k-3) + Q(k-4)\nonumber \\
&&+ Q(k-5) + Q(k-6) + Q(k-7) + Q(k-8) + Q(k-9)\nonumber \\
&&+ Q(k-10) + Q(k-13) + 2Q(k-14) + 3Q(k-15) + 4Q(k-16)\nonumber \\
&&+ 5Q(k-17) + 6Q(k-18) + 7Q(k-19) + 8Q(k-20) + 9Q(k-21)\nonumber \\
&&- 9Q(k-24) - 16Q(k-25) - 21Q(k-26) -24Q(k-27)\nonumber \\
&&- 25Q(k-28) - 24Q(k-29) - 21Q(k-30) - 16Q(k-31)\nonumber \\
&&- 9Q(k-32) - 8Q(k-36) - 21Q(k-37) - 36Q(k-38)\nonumber \\
&&- 50Q(k-39) - 60Q(k-40) - 63Q(k-41) - 56Q(k-42)\nonumber \\
&&- 36Q(k-43) + 36Q(k-47) + 84Q(k-48) + 126Q(k-49)\nonumber \\
&&+ 150Q(k-50) + 150Q(k-51) + 126Q(k-52) + 84Q(k-53)\nonumber \\
&&+ 36Q(k-54) + 28Q(k-59) + 84Q(k-60) + 150Q(k-61)\nonumber \\
&&+ 200Q(k-62) + 210Q(k-63) + 168Q(k-64) + 84Q(k-65)\nonumber \\
&&- 84Q(k-70) - 224Q(k-71) - 350Q(k-72) - 400Q(k-73)\nonumber \\
&&- 350Q(k-74) - 224Q(k-75) - 84Q(k-76) - 56Q(k-82)\nonumber \\
&&- 175Q(k-83) - 300Q(k-84) - 350Q(k-85) - 280Q(k-86)\nonumber \\
&&- 126Q(k-87) + 126Q(k-93) + 350Q(k-94) + 525Q(k-95)\nonumber \\
&&+ 525Q(k-96) + 350Q(k-97) + 126Q(k-98) + 70Q(k-105)\nonumber \\
&&+ 210Q(k-106) + 315Q(k-107) + 280Q(k-108) + 126Q(k-109)\nonumber \\
&&- 126Q(k-116) - 336Q(k-117) - 441Q(k-118) - 336Q(k-119)\nonumber \\
&&- 126Q(k-120) - 56Q(k-128) - 147Q(k-129) - 168Q(k-130)\nonumber \\
&&- 84Q(k-131) + 84Q(k-139) + 196Q(k-140) + 196Q(k-141)\nonumber \\
&&+ 84Q(k-142) + 28Q(k-151) + 56Q(k-152) + 36Q(k-153)\nonumber \\
&&- 36Q(k-162) - 64Q(k-163) - 36Q(k-164) - 8Q(k-174)\nonumber \\
&&- 9Q(k-175) + 9Q(k-185) + 9Q(k-186) + Q(k-197)\nonumber \\
&&- Q(k-208),
\label{req1}
\end{eqnarray}
which in turn gives the following equation for the topological entropy:
\begin{eqnarray}
&&x^{209} - x^{207} - x^{206} - x^{205} - x^{204} - x^{203} - x^{202}
- x^{201} - x^{200}\nonumber \\
&&- x^{199} - x^{198} - x^{195} - 2x^{194} - 3x^{193} - 4x^{192}
- 5x^{191} - 6x^{190}\nonumber \\
&&- 7x^{189} - 8x^{188} - 9x^{187} + 9x^{184} + 16x^{183}
+ 21x^{182} + 24x^{181}\nonumber \\
&&+ 25x^{180} + 24x^{179} + 21x^{178} + 16x^{177} + 9x^{176} + 8x^{172}
+ 21x^{171} \nonumber \\
&&+ 36x^{170} + 50x^{169} + 60x^{168} + 63x^{167} + 56x^{166} +
36x^{165} - 36x^{161}\nonumber \\
&&- 84x^{160} - 126x^{159} - 150x^{158} - 150x^{157} - 126x^{156}
- 84x^{155}\nonumber \\
&&- 36x^{154} - 28x^{149} - 84x^{148} - 150x^{147} - 200x^{146}
- 210x^{145} \nonumber \\
&&- 168x^{144} - 84x^{143} + 84x^{142} + 224x^{137} + 350x^{136}
+ 400x^{135}\nonumber \\
&&+ 350x^{134} + 224x^{133} + 84x^{132} + 56x^{126} + 175x^{125}
+ 300x^{124}\nonumber \\
&&+ 350x^{123} + 280x^{122} + 126x^{121} - 126x^{115} - 350x^{114}
- 525x^{113}\nonumber \\
&&- 525x^{112} - 350x^{111} - 126x^{110} - 70x^{103} - 210x^{102}
- 315x^{101}\nonumber \\
&&- 280x^{100} - 126x^{99} + 126x^{92} + 336x^{91} + 441x^{90}
+ 336x^{89}\nonumber \\
&&+ 126x^{88} + 56x^{80} + 147x^{79} + 168x^{78} + 84x^{77}
- 84x^{69}\nonumber \\
&&- 196x^{68} - 196x^{67} - 84x^{66} - 28x^{57} - 56x^{56}
- 36x^{55}\nonumber \\
&&+ 36x^{46} + 64x^{45} + 36x^{44} + 8x^{34} + 9x^{33}
- 9x^{23}\nonumber \\
&&- 9x^{22} - x^{11} + 1 = 0.
\label{equation3}
\end{eqnarray}
Resolving numerically Eq. (\ref{equation3}) one can find the biggest root
which is equal to
$$
\lambda \approx 1.61771.
$$
Correspondingly the topological entropy is given by the logarithm of
the biggest root.
We remark that the degree $(209)$
of
our equation is much smaller
than the order
of the topological
transition matrix ($2^{22}$).
It means that
most of the eigenvalues
of the topological transition matrix
are equal to zero
and, therefore,
this matrix contains much
useless information about our dynamics.
Thus we see that
the method of recurrent
relations is more economic
in this case.

One can easily see that similar calculations can be done for every
value of $n_{C}$. Here we can give the list of numbers $\lambda$
corresponding to different values of $n_{C}$:
$$
n_{C} = 2;\ \ \ \lambda \approx 1.37747
$$
$$
n_{C} = 3;\ \ \ \lambda \approx 1.51714
$$
$$
n_{C} = 4;\ \ \ \lambda \approx 1.57388
$$
$$
n_{C} = 5;\ \  \ \lambda \approx 1.59837
$$
$$
n_{C} = 10;\ \   \ \lambda \approx 1.61771
$$
$$
n_{C} = \infty;\ \ \ \lambda = (1+\sqrt{5})/2 \approx 1.61803.
$$
Thus, the topological entropy tends to the golden mean rather rapidly.

Using this algorithm it is possible to get the equations for
the topological entropy in the case of the two types of bounce (see
the end of Sec.2).
We will not present here the explicit forms of these equations, which
differ from the previous case only in the values of nonzero
coefficients, but present only some results.

First, the analog of the golden mean will be the number $2$. We can see
that there is not a smooth transition in the limit $D \to 0$ (when $D=0$
we have the simplest case with only one type of bounce). But from a
physical  point of view, when $D \to 0$ the values of the scalar field,
corresponding to the upper bounce grow, cross the Planck limit, and the
upper bounce becomes impossible. So the transition to the simplest case
takes place.

And finally, we present for illustration some values for the topological
entropy in  the symmetric case with two types of bounce:
$$
n_{C}=2;\ \ \ \lambda \approx 1.82657
$$
$$
n_{C}=3;\ \ \ \lambda \approx 1.94561
$$
$$
n_{C}=4;\ \ \ \lambda \approx 1.98273
$$
$$
n_{C}=10;\ \ \ \lambda \approx 1.99999
$$
$$
n_{C} = \infty;\ \ \ \lambda = 2.
$$

We investigate the behavior of
the topological entropy
of some cosmological
models, when they are close
to the simple isotropical closed
Friedmann-Robertson-Walker model
[its parameters (negative
cosmological constant,
charge) tend to zero].
It was shown that
chaotic behavior
is robust and topological entropy
converges to that
of the FRW model.
In the case of the
presence of
a negative cosmological constant,
we give a full description of the
corresponding symbolic dynamics
and the laws of the convergence of basic parameters in the distribution
of periodic orbits to that of the FRW model.  The situation of
the presence of matter in the model with a real scalar field and the
model with a complex scalar field are more difficult, but using the
method of recurrent relations we may find the values of the topological
entropy. This is a case when the topological entropy
is calculated for such complicated symbolic dynamics (the corresponding
topological transition matrix has 209 nonzero eigenvalues).  The
scheme described in this example can be applied to many different
physical models, obeying the different sets of prohibition rules
governing the structure of periodical trajectories.

\section*{ACKNOWLEDGMENTS}
This work was supported by Russian Foundation for Basic Research via
grants 96-02-16220 and 96-02-17591, via grant of support
of leading scientific schools 96-15-96458
and via RFBR-INTAS grant 95-1353. A.K. is grateful to CARIPLO
Scientific Foundation for the financial support during his stay in Como
in the spring of 1998.  A.K. is grateful to A.G. Kulakov for useful
conversation.

\end{document}